\documentclass[12pt]{JHEP3}

\usepackage{yfonts,amsmath,amssymb}

\newcommand{\be}{\begin{equation}}
\newcommand{\ee}{\end{equation}}
\newcommand{\beqa}{\begin{eqnarray}}
\newcommand{\eeqa}{\end{eqnarray}}

\def\im{\mathop{\rm Im}}

\newcommand{\kn}{\textswab{K}}
\relax

\def\NN{\mathcal{N}}

\def\SS{\mathcal{S}_{DBI}}
\def\dd{\mathrm{d}}
\def\dA{\mathcal{A}}
\def\TwoDers{\mathcal{T}}

\title{Holographic operator mixing - the Hall experiment}

\author{Javier Tarr\'\i o\footnote{tarrio@fpaxp1.usc.es}\\
 Departamento de F\'\i sica de Part\'\i culas,
Universidade de Santiago de Compostela \\ 
and\\
Instituto Galego de F\'\i sica de Altas Enerx\'\i as (IGFAE)\\

E-15782 Santiago
de Compostela, Spain\\
}

\abstract{We calculate the Ohmic and Hall conductivities from linear response theory in a system consisting of the intersection of $N_c$ $Dp$-branes and $N_f\ll N_c$ $Dq$-branes. Agreement is found between previous results found in a macroscopic approach comparing induced currents with external electric fields. The issue of how to deal with mixed operators is raised. The retarded Green's function is given by a matrix that can be computed from the boundary action similarly to the Green's function for uncoupled operators.}

\begin{document}

\section{Introduction}

With the proposal of the AdS/CFT conjecture \cite{hep-th/9905111} an outburst of activity in the relationship between gauge field theories and gravitational theories has appeared. The number of topics in which this gauge/gravity correspondence is shedding some light on is increasing significantly, one of the most important being the clarification of non-perturbative QCD-like theories \cite{arXiv:0901.2534}.

This regime of QCD is important to understand the phenomenology in the experiments currently carried at RHIC and shortly at LHC. Until the appearance of the AdS/CFT conjecture the only tools that could give some information about non-perturbative QCD were mainly the sum rules, chiral perturbation theory and lattice. This last lacks of a way to include real time dynamics and is not tractable analytically.

The gravity dual of QCD is not known yet, but some advances have been made to approximate it. The better understood example (and also the original one) of the duality gives the description of $\NN=4$ supersymmetric Yang-Mills \cite{hep-th/9711200}, a theory that only has matter in the adjoint representation (which we call gluons).

Karch and Katz found a way to include fundamental matter (quarks and their $\NN=2$ superpartners) in the so-called quenched approximation \cite{hep-th/0205236}, in which the number of flavor (fundamental) degrees of freedom is much less than the number of color (adjoint) degrees of freedom. This procedure does not consider backreaction of the quarks, thus corresponding to non-dynamical flavour.

The work in \cite{hep-th/0611099,hep-th/0611021} considered systems where $N_f$ $Dq$-branes were embedded in the background  of $N_c\gg N_f$ $Dp$-branes in the presence of finite baryon density, modelled by a non-trivial profile for the temporal component of the $U(1)$ gauge field in the worldvolume of the probe $Dq$-branes. Dealing with such systems one frequently finds that the equations of motion are coupled under certain generic assumptions. For example, one can consider fluctuations moving with finite momentum  and study the longitudinal component of  $U(1)$ gauge field.  At finite baryon density this channel is coupled with the scalar channel, which describes the fluctuations of the embedding profile of the probe $Dq$-branes in the $Dp$/$Dq$ system \cite{arXiv:0805.2601}. If instead of a $U(1)$ baryonic gauge group one considers a $SU(2)$ isospin group one finds coupled equations in the transverse vector channel (fluctuations of the gauge field in the directions orthogonal to the fluctuation) \cite{arXiv:0903.1864}.

This coupling of the fields in the bulk of the gravitational theory translates via the AdS/CFT correspondence to a mixing of the dual operator on the boundary gauge field theory. A clear understanding of how to treat these mixing operators is essential in order to approach QCD phenomenology from a gravity dual, and only recently this subject has been studied in depth \cite{arXiv:0903.2209,newpaper}.

In the present paper we will deal with an analytic example with an already known solution in order to compare the proposed method of dealing with operator mixing. We will calculate the Ohmic and Hall conductivities in the presence of an external magnetic field for a plasma of gluons and quarks given by a $Dp$/$Dq$ intersection. The solution was obtained in a macroscopic setup in \cite{arXiv:0708.1994}. The present calculation differs from that because we will exploit the Kubo formula. Thus we deal with linear response theory and will not need an external electric field.  Similar calculations have been done in specific cases in \cite{arXiv:0904.4772,arXiv:0704.1160,arXiv:0905.4538}.

This paper is organized as follows: in section \ref{setup} we describe the gravitational setup we are going to use. In section \ref{mixops} we define the retarded Green's function from the on-shell boundary action for the case of mixed operators. In section \ref{flucs} we calculate the Ohmic and Hall conductivities and finish in section \ref{conclusions} with some conclusions.

\section{Gravitational setup}\label{setup}
       
In this paper we will work in a general setup, but having in mind the embedding of $N_f$ probe $Dq$-branes in the background of a stack of $N_c\gg N_f$ coincident $Dp$-branes, where comparison with previous work is possible. In these systems there is a global $U(N_f)$ symmetry on the $Dq$-branes whose abelian center can be identified with a baryonic $U(1)$ symmetry, which we will exploit in order to study the system at finite baryon density. A recent review on the study of such systems can be found in \cite{arXiv:0711.4467}. In this paper we will work with a  metric of the form
\be\label{dpdqmetric}
ds^2=g_{00} (r) {\rm d}x_0^2 + g_{ii} (r) {\rm d} \vec{x}_p^2 + g_{rr} (r) {\rm d}r^2 + g_{\theta\theta} (r) {\rm d}\Omega_n^2 .
\ee
This metric is adapted to the pullback of a background of $Dp$-branes into the worldvolume of $Dq$-branes in the quenched approximation. In that case the embedding profile of the probe $Dq$-branes in the transverse space is specified by a field $\psi(r)$ which enters in (\ref{dpdqmetric}) as
$$
g_{rr}(r) = G_{rr}^{(10)}(r)+ \psi'(r)^2 G_{\psi\psi}^{(10)}(r,\psi(r)), ~~;~~ g_{\theta\theta}(r) = G^{(10)}_{\theta\theta}(r,\psi(r)),
$$
with $G^{(10)}_{ab}$ the components of the original $10$-dimensional metric given by the stack of $Dp$-branes, $\dd \Omega_n^2$ is the metric of a unit $n$-sphere and we have minkowskian signature (\emph{i.e.}, $g_{00}(r)\leq0$). Throughout this paper we will call the first three spatial coordinates\footnote{notice that hereafter it is assumed that $p>2$.}
$\{x_1, x_2, x_3,\cdots\} = \{x,y,z,\cdots\}$. Our results are generalized easily to more general setups, for example the solution of a stack of $Dp$-branes without probes.
 
 We are dealing with finite baryon density so we are forced to assume the existence of a horizon at a radius $r_H$ \cite{hep-th/0611099}. We will assume that the $g_{00}$ component of the metric vanishes linearly on the horizon as $r\to r_H$ (this can always be fulfilled by an appropriate redefinition of the radial variable). In order to have a well defined temperature one can easily show that the radial component has to diverge with the inverse of the distance to the horizon
 $g_{rr}(r\approx r_H)\approx \frac{\gamma_r}{r-r_H}$
with $\gamma_r$ a constant. Our results are invariant under radial redefinitions and so the specific behaviour of the metric near the horizon is just a matter of convenience.

 In addition we will consider a background magnetic field, consequently we have a field strength tensor of the form
\be
\frac{F_{ab}}{2\pi\alpha'}=  \left( \delta_{ar} \delta_{b0} - \delta_{a0} \delta_{br}  \right)  A_t'(r) +  \left( \delta_{ay} \delta_{bx} - \delta_{ax} \delta_{by}  \right)  B_z.
\ee
The constant $B_z$ we will call a magnetic field, even when $p>3$.

Having in mind the DBI action, it is useful to define the matrix 
\be \label{gammamatrix}
 \gamma_{ab}(r)=g_{ab}(r)+F_{ab}(r) 
\ee
which is a function only of the radial variable appearing in the metric (\ref{dpdqmetric}) and has only diagonal componets apart from the two antisymmetric components coming from the electromagnetic tensor $\gamma_{0r}=-\gamma_{r0}$ and $\gamma_{xy}=-\gamma_{yx}$.

Notice also that the $\gamma_{ii}$ factors are all the same, but this is not the true for the inverse of the $\gamma_{ab}$ matrix
\beqa
&&\gamma_{00}=-\gamma_{ii} f(r) ~~;~~ \gamma_{rr}=G(r) f(r)^{-1} ~~;~~ f(r)=(r-r_H) \kn(r)\, , \nonumber\\
&&\gamma^{00}=\frac{\gamma_{rr}}{\gamma_{00}\gamma_{rr}+(\gamma_{0r})^2} ~~;~~\gamma^{\perp\perp}=\frac{\gamma_{ii}}{(\gamma_{ii})^2+(\gamma_{xy})^2}\, ~~;~~\gamma^{zz}=\frac{1}{\gamma_{ii}}, \nonumber \\
&&\gamma^{rr}= \frac{\gamma_{00}}{\gamma_{00}\gamma_{rr}+(\gamma_{0r})^2}~~;~~\gamma^{0r}= -\gamma^{r0}=\frac{-\gamma_{0r}}{\gamma_{00}\gamma_{rr}+(\gamma_{0r})^2}, \nonumber\\
&&\gamma^{\theta\theta}=\frac{1}{\gamma_{\theta\theta}}~~;~~\gamma^{xy}= -\gamma^{yx}=\frac{-\gamma_{xy}}{(\gamma_{ii})^2+(\gamma_{xy})^2} \nonumber \, .
\eeqa
where $\gamma_{ii}$, $\gamma_{\theta\theta}$, $\gamma_{0r}$, $\gamma_{xy}$, $\kn(r)$ and $G(r)$ are regular at the horizon  and $\gamma^{ab}\gamma_{ca}=\delta^b_c$. $\gamma^{\perp\perp}$ stands for $\gamma^{xx}=\gamma^{yy}$ and $\gamma^{zz}$ for all the other $p-2$ flat-spatial diagonal components of the $\gamma$ matrix.

\subsection{Background field solution}

To write a solution for the background gauge field components we will be focusing on the DBI action
\be\label{origdbi}
\SS=-N_f T_{D_q}  \int \dd^p x\, \dd r\, \dd \Omega_n\, e^{-\phi} \sqrt{-\gamma} ,
\ee
where $\gamma\equiv \det(\gamma_{ab}) = g_{ii}^{p-2 }g_{\theta\theta}^n \left(g_{00}g_{rr}+(2\pi\alpha')^2 A_t'^2\right)  \left( (g_{ii})^2+(2\pi\alpha')^2 B_z^2 \right)$ and $\dd \Omega_n$ is the volume form of a unit $n$-sphere. Notice that we are not considering Wess Zumino terms, which are not needed in the present case. Its presence is discussed further in \cite{arXiv:0708.1994,arXiv:0905.4538,arXiv:0809.3808}. The equation of motion one reads from (\ref{origdbi}) for the background gauge field is
\be\label{bggaugefield}
\frac{\delta \SS}{\delta A_t'} = N_f T_{D_q} (2\pi\alpha') e^{-\phi}\sqrt{-\gamma}\gamma^{0r} = \frac{n_q}{\Omega_n},
\ee
with $\Omega_n$ the volume of the unit $n$-sphere. As all the angular dependence is encoded in the volume form, we can identify the constant $n_q$ with the density of quarks in the field theory \cite{hep-th/0611099}. We can use this constant of motion $n_q$ to find a solution for the gauge field component
\be
A_t'(r)  =  n_q \sqrt{\frac{-g_{00} g_{rr}}{\NN^2 e^{-2\phi}  g_{ii}^{p-2} \left( g_{ii}^2 + (2\pi \alpha')^2 B_z^2 \right) g_{\theta\theta}^n +  (2\pi\alpha')^2 n_q^2} }, \label{bgeqA0}
\ee
with $\NN= N_f T_{D_q} \Omega_n (2\pi\alpha')^2$. The holographic dictionary tells us that the value of this field at the boundary is related to the chemical potential, $\mu$, for the quarks, whereas at the horizon one has to impose vanishing of the gauge field component to avoid having a singular one-form \cite{hep-th/0611099}, thus we find for the chemical potential
$$
\mu = n_q  \int_{r_H}^{r_{bou}} \sqrt{\frac{-g_{00} g_{rr}}{\NN^2 e^{-2\phi}  g_{ii}^{p-2} \left( g_{ii}^2 + (2\pi \alpha')^2 B_z^2 \right) g_{\theta\theta}^n +  (2\pi\alpha')^2 n_q^2} }\dd r.
$$

For the embedding profile of the probe $D7$-branes $\psi(r)$, the equation of motion one finds is a highly non-linear one. The solutions depend non-analytically on two parameters $\psi_0$ (the angle at which the branes enter the black hole) and $n_q$ \cite{hep-th/0611099}.

\section{Holographic retarded correlator for mixed operators}\label{mixops}

Mixed operators arise naturally in holographic models in the form of coupled differential equations for the fluctuations of the background fields. On occasions these couplings involve non-physical fields, as is the case of the temporal and longitudinal fluctuations of a one-form, which can be combined in terms of a gauge invariant quantity to decouple the system \cite{hep-th/0506184}. On the other hand, at times the coupled equations appearing already involve physical gauge invariant fields. When this occurs the operators get mixed in the IR of the gauge theory. This mixing can be propagated to the UV.

There are several examples in the literature where one finds this mixing of operators. In particular it appears in the study of the $D3$/$D7$ system at finite baryon density as a coupling between the longitudinal and scalar perturbations \cite{arXiv:0805.2601}. A similar system, but with finite isospin density instead of finite baryon density, also has such a coupling in the transverse channel, giving rise to a superfluid behaviour of the system \cite{arXiv:0903.1864}. These issues present some subtleties on the study of these systems, for example, if one wants to study the quasinormal modes of a system with operator mixing one has be particularly careful, as shown for example in \cite{arXiv:0903.2209, newpaper}.

We will bring here a simplified discussion of the holographic prescription to study this particular kind of systems. The linearized equations of motion we want to solve can be derived from the action\footnote{note that this action for the fluctuations is just a Maxwell-like one, so this discussion is independent of the fact that it originally comes from a DBI action.}
\be\label{genaction}
S  =  -\frac{K}{2} \int\dd^p x\, \dd r\,  e^{-\phi} \sqrt{-\gamma}\, \gamma^{\mu\nu}  \partial_\mu \tilde\Phi^T \cdot \TwoDers \cdot \partial_\nu \tilde\Phi\,,
\ee
where $\TwoDers$ is a matrix which couples the different fields (given by the column vector $\tilde\Phi(t,x_p,r)$) in our bulk theory. The form of this matrix is responsible for the mixing of operators at the boundary.

Assuming the $\gamma$ components are of the form (\ref{gammamatrix}), the Euler-Lagrange equations after expressing the fields in the mode representation $\tilde \Phi(t,x_p,r)=e^{-i(\omega t - q x_p)}\Phi(r)$ read
 \be\label{mageoms}
\partial_r \left( e^{-\phi} \sqrt{-\gamma} \gamma^{rr} \TwoDers_S \cdot \Phi'  \right)+ \left[ i \omega \partial_r \left( e^{-\phi} \sqrt{-\gamma} \gamma^{0r} \TwoDers_A \right) - e^{-\phi}\sqrt{-\gamma} \left( \omega^2 \gamma^{00} + q^2 \gamma^{zz} \right) \TwoDers_S  \right] \cdot \Phi =0,
\ee
with $\TwoDers_{S(A)}$ the (anti)symmetric part\footnote{we follow the convention $\TwoDers_S=\frac{\TwoDers+\TwoDers^T}{2}$ and similarly for the antysymmetric part.} of  $\TwoDers$.

One can express the on-shell boundary action from (\ref{genaction}) as
\be
S_{bou} =  -\frac{K}{2} \int\dd^p x\, e^{-\phi} \sqrt{-\gamma}\,  \tilde\Phi^T \cdot \left[ \gamma^{rr} \TwoDers_S \cdot \partial_r \tilde\Phi\,-\gamma^{0r} \TwoDers_A \cdot \partial_0 \tilde\Phi\, \right]_{r_H}^{\Lambda}  ,
\ee
where we evaluate it at a cutoff $r=\Lambda$. Reexpressing it as the product of a profile and a source $\tilde \Phi(t,x_p,r)=e^{-i(\omega t - q x_p)}F_k(r) \cdot \varphi(k)$ with $F_k(\Lambda) = {\mathbf{I}}$; the Fourier transformed action reads
\beqa
S_{bou} & = &  -\frac{K}{2} \int \frac{\dd^p k}{(2\pi)^d}\, e^{-\phi} \sqrt{-\gamma}\,  \varphi^T(-k) \cdot \left[ \gamma^{rr}  F_{-k}^T \cdot  \TwoDers_S \cdot F_k' +i \omega \gamma^{0r} F_{-k}^T \cdot  \TwoDers_A \cdot F_k \right] \cdot \varphi(k)  \nonumber \\
& \equiv & \frac{1}{2} \int \frac{\dd^p k}{(2\pi)^p}  \varphi^T(-k) \cdot {\cal F}_k\cdot \varphi(k)\Big|^{\Lambda}_{r_H}.
\eeqa

The gauge/gravity conjecture relates this on-shell boundary action with the retarded Green's function by
$$
G^R (k)= - \lim_{\Lambda\to\infty} {\cal F}_k(\Lambda),
$$
which for the case of unmixed operators (\emph{i.e.}, a diagonal $\TwoDers$) coincides with the prescription given in \cite{hep-th/0205051}, which has been derived also via the Schwinger-Keldish formalism \cite{hep-th/0212072,arXiv:0902.4010}. One can show that $G^R(-k)=G^R(k)^*=G^A(k)$, see appendix A of \cite{newpaper}.

When dealing with unmixed operators it is known that one can define a conserved flux
$$
\partial_r \im G^R(k) = 0.
$$
This is promoted to the condition
\be
\partial_r \left(  {\cal F}_k -  {\cal F}_k^* \right) = F_{-k}^T \cdot \partial_r \left(  e^{-\phi} \sqrt{-\gamma}\, \gamma^{rr} \TwoDers_S \cdot F_k' \right) - \partial _r \left( e^{-\phi} \sqrt{-\gamma}\, \gamma^{rr} \left(F_{-k}^T\right)' \cdot \TwoDers_S \right)\cdot F_k = 0\, ,\label{flux}
\ee
where we used ${\cal F}_k^* = {\cal F}_{-k}$ and  the last equality is guaranteed by the equations of motion. This conserved flux can also be seen as the conservation of a Noether current due to a $U(1)$ symmetry of the bilinear action \cite{newpaper}.

\section{Fluctuations and the conductivity matrix}\label{flucs}

We will now introduce perturbations to our system. Given the existence of a non vanishing background field $B_z$ the most general perturbation we can write is expressed as a mode $e^{ -i (\omega t -q_x x-q_z z) }$. Having the Hall effect in mind we want to make perturbations of the electric field in the directions perpendicular to the background magnetic field. For this purpose we will restrict to the simpler case of a fluctuation parallel to the background magnetic field ($q_x=0$), in which the possible perturbations appearing can be classified according to their transformation under the little group $SO(p-1)$ \cite{hep-th/0506184}.

The Kubo formula we will use to calculate the conductivity matrix requires the evaluation of the transverse channel given by the perturbations $\dA_{p\neq z}(t,z,r)$. The presence of a magnetic field given by a gauge potential $A_x(y)$ couples the fields
$\dA_x$ and $\dA_y$, leaving the other transverse components independent. Including the perturbations, we shall deal with a gauge field of the following form
\be
A_\mu = (A_t(r), B_z y  + \dA_x(t,z,r) , \dA_y(t,z,r), 0, \cdots, 0)
\ee
which amounts to doing perturbations on top of a macroscopic magnetic field. One can find the solution for the other transverse perturbations $\dA_{p\neq \{x,y,z\}}$ from these two by taking the $B_z\to0$ limit.

To study the linearized equations of motion one should expand the DBI action to second order in fluctuations. One can do this and then compare to the general expression (\ref{genaction}), finding
\be\label{coefs}
\TwoDers  =  \begin{pmatrix}
\gamma^{\perp\perp} & \gamma^{xy} \\
 -\gamma^{xy} & \gamma^{\perp\perp}
\end{pmatrix}   ~~;~~ \Phi = \begin{pmatrix}
\dA_x \\
\dA_y
\end{pmatrix} ~~;~~ K= \NN= N_f T_{D_q} \Omega_n (2\pi\alpha')^2.
\ee

These equations can be solved in the hydrodynamic limit\footnote{in this limit the equations decouple in the appropriate basis, but this fact does not affect our discussion on operator mixing.}, as shown in the appendix. The final solution for $\Phi$ in this limit is
\beqa
\Phi(r) & = & \left[  {\mathbf{I}} - i\, \omega \int_{C_2}^{r} \frac{{\mathcal{C}}}{e^{-\phi}\sqrt{-\gamma}\gamma^{\perp\perp}\gamma^{rr}} - i\, \omega \int_{C_2}^{r} \frac{\gamma^{xy}\gamma^{0r}}{\gamma^{\perp\perp}\gamma^{rr}} \begin{pmatrix}
0 & -1 \\ 1 & 0
\end{pmatrix} +{\mathcal{O}}(\omega^2, q_z^2) \right] \cdot \varphi (k) \nonumber \\
 & = & F_k(r) \cdot \varphi (k),\label{Bfieldsols}
\eeqa
where $C_2$ is chosen such that $F_k(\Lambda) = {\mathbf{I}}$, $\varphi (k) = \Phi(\Lambda)$ and ${\mathcal{C}}$ is the constant matrix
\be \label{Bfieldsols2}
{\mathcal{C}} = e^{-\phi}\sqrt{-\gamma} \begin{pmatrix}
\sqrt{-\gamma^{00}\gamma^{rr}}\gamma^{\perp\perp} & \gamma^{xy}\gamma^{0r} \\ -\gamma^{xy}\gamma^{0r} & \sqrt{-\gamma^{00}\gamma^{rr}}\gamma^{\perp\perp}
\end{pmatrix}\Bigg|_{r\to r_H}.
\ee
It can be shown that the antihermitian part of the flux (\ref{flux}) is conserved at the given order in $\omega$ and $q_z$.

Following the formalism described in section \ref{mixops} we can read off the retarded transverse Green's function as
$$
G^R_\perp (k)= - \lim_{\Lambda\to\infty} {\cal F}_k(\Lambda) = \NN \lim_{r \to\infty}e^{-\phi} \sqrt{-\gamma}\, \left[  \TwoDers_S \cdot F_k' +i \omega \gamma^{0r}  \TwoDers_A \right] = -i\, \omega\, \NN\,  {\mathcal{C}} + \cdots \,,
$$
where the dots express higher order in the frequency and momentrum. From the former equation we can evaluate the matrix of conductivities as
\be\label{maxconducs}
\sigma(\omega) = \frac{i\, G^R_\perp(\omega = |q_z|)}{\omega} = \NN {\mathcal{C}} + {\mathcal{O}}(\omega).
\ee

As we are working in the hydrodynamic regime we can only study the DC conductivity $\sigma(0)$, given by 
\be\label{result}
\sigma(0) = \begin{pmatrix} \sigma_{xx} & \sigma_{xy} \\ -\sigma_{xy} & \sigma_{xx} \end{pmatrix},
\ee
where
\be\label{resultdets}
\sigma_{xx} = \NN e^{-\phi} \sqrt{\gamma \gamma^{00} \gamma^{rr}} \gamma^{\perp\perp} \Big|_{r_H},   ~~~~  \sigma_{xy} = \NN e^{-\phi} \sqrt{-\gamma} \gamma^{xy} \gamma^{0r} \Big|_{r_H}  .
\ee

This result depends only on the bilinear action (\ref{genaction}) with coefficients (\ref{coefs}) and the presence of a suitable black hole, so this formula can apply with minor modifications to other setups apart from $Dp$/$Dq$ intersections. Due to the linearization of the perturbations the importance of the original DBI action in these results gets reduced.

Additional transverse components (appearing when $p>3$) will give a diagonal conductivity given by the $B_z\to 0$ limit of (\ref{resultdets}), \emph{i.e.}
$$
\sigma = \NN e^{-\phi} \sqrt{\gamma_{\left( B_z\to0\right)} \gamma^{00} \gamma^{rr}} \gamma^{zz} \Big|_{r_H}.
$$
This is the result obtained in \cite{arXiv:0809.3808,arXiv:0811.1750}.

When the dependence on the charge density is explicitly needed one has to plug a solution for the background gauge field $A_t$. In this case the form of the original action becomes relevant. For example, plugging the results (\ref{bggaugefield})  and  (\ref{bgeqA0}) in (\ref{resultdets}) one obtains for the DC conductivity and the Hall coefficient as originally found in \cite{arXiv:0708.1994}
\be\label{obannonres}
\sigma_{xx} = \sqrt{\frac{\NN^2 e^{-2\phi} \gamma_{ii}^{p} \gamma_{\theta\theta}^n}{ \gamma_{ii}^2 + (2\pi\alpha')^2 B_z^2} + \frac{(2\pi\alpha')^2 n_q^2 \gamma_{ii}^2}{ \left(\gamma_{ii}^2 + (2\pi\alpha')^2 B_z^2 \right)^2}} \Bigg|_{r_H}, ~~~ \sigma_{xy} = \frac{(2\pi\alpha')^2 n_q B_z}{(\gamma_{ii}(r_H))^2+(2\pi\alpha')^2 B_z^2}.
\ee

The insertion of a background electric field in our setup would have complicated the analysis performed. An external electric field induces the presence of a shell outside the horizon (described by a radius related to the modulus of the electric field) on which all the fields must be regularized. In a black hole setup this induces a current on the boundary that can be analyzed to get the values for the conductivity and the Hall coefficient, whose zero electric field limit is given by (\ref{obannonres}).

As noted in \cite{arXiv:0708.1994,arXiv:0905.4538,arXiv:0809.3808}, adding a $\theta$ term to the original action has as an effect on the present result a shift proportional to $\theta$ in the charge density and in the Hall coefficient.

\section{Conclusions}\label{conclusions}

We used a generalized prescription to obtain the Green's function in the case of operators mixing on the boundary to calculate the conductivity and the Hall coefficient of a plasma with a gravity dual given by a system of $N_f\ll N_c$ $Dq$-branes in the background of $N_c$ $Dp$-branes. The solution to this problem  in a macroscopic setup was found in \cite{arXiv:0708.1994,arXiv:0705.3870}, where the matrix of conductivities was computed via the relation between induced currents on the boundary and external electric fields
$$
j_i = \sigma_{ji}E_j.
$$
The present calculation complements the previous, macroscopic approach with the microscopic procedure, making use of the Kubo formula (\ref{maxconducs}). Agreement between both procedures is found.

Both methods are based in the regularisation of the fields at a given radius, set by the external electric field. When this electric field is set to zero the radius at which one has to regularise coincides with the radius of the horizon. The transport phenomena information one can extract from this kind of systems is encoded uniquely by this surface, and it appears naturally by imposing regular behaviour.

An important point given in \cite{arXiv:0708.1994} is that the expressions  (\ref{obannonres}), evaluated at the radius $r_\star$ instead of the horizon, would give the result for the conductivity and the Hall coefficient in the presence of an electric field $E$ creating a singular shell at precisely that radius $r_\star(E)$. We suggest that this still holds in the more general result (\ref{resultdets}). In \cite{arXiv:0904.3905} the conductivity was calculated in a setup with just an external electric field and this suggestion was found to be true.

\acknowledgments

I would like to thank Karl Landsteiner, Javier Mas and Jonathan Shock for useful comments and discussions. This work was supported in part by the MEPSYD and FEDER (grant FPA200801838), by the Spanish ConsoliderIngenio 2010 Programme CPAN (CSD200700042), by Xunta de Galicia (Conselleria de Educacion and grant PGIDIT06 PXIB206185PR) and by MEPSYD of Spain under a grant of the FPU program.

\appendix

\section{Solving the equations of motion}

The equations of motion (\ref{mageoms}) can be written as
\beqa
 \Phi'' &&+ \log'\left[  e^{-\phi}\sqrt{-\gamma} \gamma^{\perp\perp}\gamma^{rr} \right] \Phi'  - \frac{ \omega^2 \gamma^{00} + q_z^2 \gamma^{zz}}{\gamma^{rr}} \Phi  \nonumber\\
&&+ \omega \log' \left[  e^{-\phi}\sqrt{-\gamma} \gamma^{xy} \gamma^{0r} \right] \frac{\gamma^{xy} \gamma^{0r}}{\gamma^{\perp\perp} \gamma^{rr}} \sigma^2 \cdot \Phi =0,
\eeqa
with $\sigma^2$ the second Pauli matrix. This can be solved by shifting to the basis of polarizations 
$$
\Psi \equiv \begin{pmatrix} \dA_+ \\ \dA_-\end{pmatrix} = \begin{pmatrix}  1 & i \\ 1 & -i\end{pmatrix}\cdot \Phi,
$$
 in which the equation becomes
\beqa
\Psi'' &&+ \log'\left[ e^{-\phi} \sqrt{-\gamma} \gamma^{\perp\perp}\gamma^{re^{-\phi}r} \right] \Psi' - \frac{ \omega^2 \gamma^{00} + q_z^2 \gamma^{zz}}{\gamma^{rr}} \Psi \nonumber\\
&&- \omega \log' \left[ e^{-\phi} \sqrt{-\gamma} \gamma^{xy} \gamma^{0r} \right] \frac{\gamma^{xy} \gamma^{0r}}{\gamma^{\perp\perp} \gamma^{rr}}  \sigma^3 \cdot \Psi ,
\eeqa
where $\sigma^3$ is the third (and thus diagonal) Pauli matrix. Thus, these are two uncoupled equations for the components of $\Psi$.

 The horizon ($r=r_H$) is a regular singular point, therefore we can perform the usual Frobenius study to find its singular behaviour. The indices at the horizon are
$$
\eta_\pm = \pm i \frac{\omega}{4 \pi T}\, ,
$$
and we will consider just the negative sign to regularize the function in that point 
$$\dA_\pm=f^{\eta_-}\dA_{\pm,reg}\, .$$

\subsection{Hydrodynamic solution}

We can make an expansion at low $\omega$ and $q$. To do this we set
\beqa
(\omega, q_z ) &\rightarrow& \lambda_{hyd} (\omega, q_z)\, , \nonumber\\ 
\dA_\pm &\rightarrow& f^{\eta_{-}}\left(\dA_\pm^{(0)} + \lambda_{hyd} \dA_\pm^{(1)}\right)= \dA_\pm^{(0)} + \lambda_{hyd} \left( \dA_\pm^{(1)} + \eta_- \dA_\perp^{(0)} \log f  \right) \, .\nonumber
\eeqa
One finds that $\dA_{\pm}^{(0)}$ is a constant and an equation for $\dA_{\pm}^{(1)}$
\be
\partial_r \left( e^{-\phi}\sqrt{-\gamma} \gamma^{ii}\gamma^{rr} \partial_r \dA^{(1)}_{\pm}\right) = 
\pm \omega \partial_r\left(e^{-\phi}\sqrt{-\gamma}\gamma^{xy}\gamma^{0r}\right) \dA_{\pm}^{(0)}\, ,
\ee
which can be solved as follows
\be
\dA^{(1)}_{\pm}(r)  =  \int_{C_2}^r \frac{C_1^{\pm} \pm \omega  e^{-\phi}\sqrt{-\gamma}\gamma^{xy}\gamma^{0r}\dA_{\pm}^{(0)}}{e^{-\phi}\sqrt{-\gamma}\gamma^{ii}\gamma^{rr}} \, ,
\ee
where the constant $C_2$ is the same for both solutions.

We can now express the solution for $\dA_{x,y}$ as
\beqa
\dA_x & = & \frac{\dA_+^{(0)}+\dA_-^{(0)}}{2} + \int_{C_2}^r \frac{\frac{C_1^{+}+C_1^{-}}{2} + \omega  e^{-\phi}\sqrt{-\gamma}\gamma^{xy}\gamma^{0r}\frac{\dA_+^{(0)}-\dA_-^{(0)}}{2}}{e^{-\phi}\sqrt{-\gamma}\gamma^{ii}\gamma^{rr}} + {\mathcal{O}}(\lambda_{hyd}^2) \, , \\
\dA_y & = & \frac{\dA_+^{(0)}-\dA_-^{(0)}}{2i} +  \int_{C_2}^r \frac{\frac{C_1^{+}-C_1^{-}}{2i} + \omega  e^{-\phi}\sqrt{-\gamma}\gamma^{xy}\gamma^{0r}\frac{\dA_+^{(0)}+\dA_-^{(0)}}{2i}}{e^{-\phi}\sqrt{-\gamma}\gamma^{ii}\gamma^{rr}}+ {\mathcal{O}}(\lambda_{hyd}^2) \, .
\eeqa

In order to fix the integration constants $\dA_\pm^{(0)}$ we normalize the fields $\dA_{x,y}$ to the value at the cut-off $r=\Lambda$. For this it suffices to take $\dA_\pm^{(0)}=\dA_x^{\Lambda}\pm i \dA_y^{\Lambda}$ and $C_2$ such that the second term in the solution vanishes at the cut-off, and from this we read
\beqa
\dA_x & = & \dA_x^{\Lambda} + \int_{C_2}^r \frac{C_x +  i \omega e^{-\phi}\sqrt{-\gamma}\gamma^{xy}\gamma^{0r} \dA_y^{\Lambda}}{e^{-\phi}\sqrt{-\gamma}\gamma^{ii}\gamma^{rr}} + {\mathcal{O}}(\lambda_{hyd}^2) \, , \\
\dA_y & = & \dA_y^{\Lambda} + \int_{C_2}^r \frac{C_y -  i \omega  e^{-\phi}\sqrt{-\gamma}\gamma^{xy}\gamma^{0r} \dA_x^{\Lambda}}{e^{-\phi}\sqrt{-\gamma}\gamma^{ii}\gamma^{rr}}+ {\mathcal{O}}(\lambda_{hyd}^2) \, .
\eeqa

The regularizing constants $C_{x}$ and $C_y$ are fixed by imposing regularity of the function $\dA^{(1)}_{x,y,reg}(r)=\dA^{(1)}_{x,y}(r)-\frac{i \omega}{4\pi T}\log f \dA_{x,y}^{\Lambda}$ on the horizon. For $\dA_x$ we obtain after derivation the condition
$$
\frac{C_x +  i \omega e^{-\phi}\sqrt{-\gamma}\gamma^{xy}\gamma^{0r}\dA_y^{\Lambda}}{e^{-\phi}\sqrt{-\gamma}\gamma^{ii}\gamma^{rr}} - \frac{i \omega}{4\pi T}\frac{f'}{f}\dA_x^{\Lambda}\Big|_{r\to r_H} = \mathrm{finite} \, ,
$$
and we finally obtain (using the fact that $e^{-\phi}\sqrt{-\gamma}\gamma^{xy}\gamma^{0r}$ is finite on the horizon)
\beqa 
C_x & = & -i \omega\, e^{-\phi}\sqrt{-\gamma} \left[Ê\sqrt{-\gamma^{00}\gamma^{rr}}\gamma^{ii} \dA_x^{\Lambda}+ \gamma^{xy}\gamma^{0r} \dA_y^{\Lambda}\right]_{r\to r_H} \, ,\\
C_y & = & -i \omega\, e^{-\phi}\sqrt{-\gamma} \left[Ê\sqrt{-\gamma^{00}\gamma^{rr}}\gamma^{ii} \dA_y^{\Lambda}- \gamma^{xy}\gamma^{0r} \dA_x^{\Lambda}\right]_{r\to r_H} \, .
\eeqa


\begin{thebibliography}{99}
\bibitem{hep-th/9905111}
  O.~Aharony, S.~S.~Gubser, J.~M.~Maldacena, H.~Ooguri and Y.~Oz,
  ``Large N field theories, string theory and gravity,''
  Phys.\ Rept.\  {\bf 323}, 183 (2000)
  [arXiv:hep-th/9905111].
  
\bibitem{arXiv:0901.2534}
  J.~D.~Edelstein, J.~P.~Shock and D.~Zoakos,
  ``The AdS/CFT Correspondence and Non-perturbative QCD,''
  AIP Conf.\ Proc.\  {\bf 1116}, 265 (2009)
  [arXiv:0901.2534 [hep-ph]].
  
\bibitem{hep-th/9711200}
  J.~M.~Maldacena,
  ``The large N limit of superconformal field theories and supergravity,''
  Adv.\ Theor.\ Math.\ Phys.\  {\bf 2}, 231 (1998)
  [Int.\ J.\ Theor.\ Phys.\  {\bf 38}, 1113 (1999)]
  [arXiv:hep-th/9711200].

\bibitem{hep-th/0205236}
  A.~Karch and E.~Katz,
  ``Adding flavor to AdS/CFT,''
  JHEP {\bf 0206}, 043 (2002)
  [arXiv:hep-th/0205236].

\bibitem{hep-th/0611099}
  S.~Kobayashi, D.~Mateos, S.~Matsuura, R.~C.~Myers and R.~M.~Thomson,
  ``Holographic phase transitions at finite baryon density,''
  JHEP {\bf 0702}, 016 (2007)
  [arXiv:hep-th/0611099].

\bibitem{hep-th/0611021}
  S.~Nakamura, Y.~Seo, S.~J.~Sin and K.~P.~Yogendran,
  ``A new phase at finite quark density from AdS/CFT,''
  J.\ Korean Phys.\ Soc.\  {\bf 52}, 1734 (2008)
  [arXiv:hep-th/0611021].

\bibitem{arXiv:0805.2601}
  J.~Mas, J.~P.~Shock, J.~Tarrio and D.~Zoakos,
  ``Holographic Spectral Functions at Finite Baryon Density,''
  JHEP {\bf 0809}, 009 (2008)
  [arXiv:0805.2601 [hep-th]].

\bibitem{arXiv:0903.1864}
  M.~Ammon, J.~Erdmenger, M.~Kaminski and P.~Kerner,
  ``Flavor Superconductivity from Gauge/Gravity Duality,''
 JHEP {\bf 0910}, 067 (2009)
  [arXiv:0903.1864 [hep-th]].
  
\bibitem{arXiv:0903.2209}
  I.~Amado, M.~Kaminski and K.~Landsteiner,
  ``Hydrodynamics of Holographic Superconductors,''
  JHEP {\bf 0905}, 021 (2009)
  [arXiv:0903.2209 [hep-th]].

\bibitem{newpaper}
  M.~Kaminski, K.~Landsteiner, J.~Mas, J.~P.~Shock and J.~Tarrio,
  ``Holographic Operator Mixing and Quasinormal Modes on the Brane,''
  arXiv:0911.3610 [hep-th].

 
\bibitem{arXiv:0708.1994}
  A.~O'Bannon,
  ``Hall Conductivity of Flavor Fields from AdS/CFT,''
  Phys.\ Rev.\  D {\bf 76}, 086007 (2007)
  [arXiv:0708.1994 [hep-th]].
  
\bibitem{arXiv:0904.4772}
  G.~Lifschytz and M.~Lippert,
  ``Anomalous conductivity in holographic QCD,''
  Phys.\ Rev.\  D {\bf 80}, 066005 (2009)
  [arXiv:0904.4772 [hep-th]].

\bibitem{arXiv:0704.1160}
  S.~A.~Hartnoll and P.~Kovtun,
  ``Hall conductivity from dyonic black holes,''
  Phys.\ Rev.\  D {\bf 76}, 066001 (2007)
  [arXiv:0704.1160 [hep-th]].

\bibitem{arXiv:0905.4538}
  J.~Alanen, E.~Keski-Vakkuri, P.~Kraus and V.~Suur-Uski,
  ``AC Transport at Holographic Quantum Hall Transitions,''
  JHEP {\bf 0911}, 014 (2009)
  [arXiv:0905.4538 [hep-th]].
  
\bibitem{arXiv:0711.4467}
  J.~Erdmenger, N.~Evans, I.~Kirsch and E.~Threlfall,
  ``Mesons in Gauge/Gravity Duals - A Review,''
  Eur.\ Phys.\ J.\  A {\bf 35}, 81 (2008)
  [arXiv:0711.4467 [hep-th]].

\bibitem{arXiv:0809.3808}
  N.~Iqbal and H.~Liu,
  ``Universality of the hydrodynamic limit in AdS/CFT and the membrane
  paradigm,''
  Phys.\ Rev.\  D {\bf 79}, 025023 (2009)
  [arXiv:0809.3808 [hep-th]].

\bibitem{hep-th/0506184}
  P.~K.~Kovtun and A.~O.~Starinets,
  ``Quasinormal modes and holography,''
  Phys.\ Rev.\  D {\bf 72}, 086009 (2005)
  [arXiv:hep-th/0506184].

\bibitem{hep-th/0205051}
  D.~T.~Son and A.~O.~Starinets,
  ``Minkowski-space correlators in AdS/CFT correspondence: Recipe and
  applications,''
  JHEP {\bf 0209}, 042 (2002)
  [arXiv:hep-th/0205051].

\bibitem{hep-th/0212072}
  C.~P.~Herzog and D.~T.~Son,
  ``Schwinger-Keldysh propagators from AdS/CFT correspondence,''
  JHEP {\bf 0303}, 046 (2003)
  [arXiv:hep-th/0212072].

\bibitem{arXiv:0902.4010}
  B.~C.~van Rees,
  ``Real-time gauge/gravity duality and ingoing boundary conditions,''
  Nucl.\ Phys.\ Proc.\ Suppl.\  {\bf 192-193}, 193 (2009)
  [arXiv:0902.4010 [hep-th]].
  
 \bibitem{arXiv:0811.1750}
  J.~Mas, J.~P.~Shock and J.~Tarrio,
  ``A note on conductivity and charge diffusion in holographic flavour
  systems,''
  JHEP {\bf 0901}, 025 (2009)
  [arXiv:0811.1750 [hep-th]].

\bibitem{arXiv:0705.3870}
  A.~Karch and A.~O'Bannon,
  ``Metallic AdS/CFT,''
  JHEP {\bf 0709}, 024 (2007)
  [arXiv:0705.3870 [hep-th]].

\bibitem{arXiv:0904.3905}
  J.~Mas, J.~P.~Shock and J.~Tarrio,
  ``Holographic Spectral Functions in Metallic AdS/CFT,''
  JHEP {\bf 0909}, 032 (2009)
  [arXiv:0904.3905 [hep-th]].



 \end{thebibliography}
\end{document}